\documentclass[12pt]{article}
\usepackage{amsthm,amsmath,amsfonts,amssymb}
\usepackage{graphicx}
\usepackage{enumerate}
\usepackage{natbib}
\usepackage{url} \usepackage{hyperref}
\usepackage{booktabs}
\usepackage{etoolbox}

\usepackage[left=0.9in,right=0.9in,top=0.95in,bottom=0.95in]{geometry}

\RequirePackage{tikz-cd}

\usepackage{pdflscape} \usepackage{rotating}
\usepackage{placeins}

\theoremstyle{plain}

\newtheorem{theorem}{Theorem}
\newtheorem{lemma}{Lemma}
\newtheorem{proposition}{Proposition}
\newtheorem{corollary}{Corollary}
\theoremstyle{remark}
\newtheorem{definition}{Definition}

\newtheorem{assumption}{Assumption}

\makeatletter
\DeclareRobustCommand*\textsubscript[1]{\@textsubscript{\selectfont#1}}
\def\@textsubscript#1{{\m@th\ensuremath{_{\mbox{\fontsize\sf@size\z@#1}}}}}
\makeatother

\begin{document}

\def\spacingset#1{\renewcommand{\baselinestretch}{#1}\small\normalsize} \spacingset{1}

\apptocmd{\align}{\vspace{-5pt}}{}{}
\pretocmd{\align}{\vspace{-5pt}}{}{}

\newcommand{\blind}{1}

\if1\blind
{
\title{\bf Causal Generalization of Continuous Treatment Effects under Covariate Shift}

\author{
Jay Jojo Cheng$^{1}$ and Guanhua Chen$^{1,*}$ \\
$^{1}$Department of Biostatistics and Medical Informatics, \\
University of Wisconsin--Madison, Madison, Wisconsin, USA \\
$^{*}$Corresponding author: \texttt{gchen25@wisc.edu}
}

\maketitle
}
\fi
\if0\blind
{
  \bigskip
  \bigskip
  \bigskip
   \title{\bf  Causal Generalization of Continuous Treatment Effects under Covariate Shift}
	\author{}
	\date{}
  \maketitle
}\fi

\bigskip
\begin{abstract}
Average dose-response functions are widely used to summarize causal
effects of continuous treatments, but most existing methods assume that
the observed sample represents the target population. We study a
covariate-shift setting in which covariates, treatment, and outcome are
observed in a labelled source sample, while only covariates are observed
in the target sample. We develop a two-sample local polynomial regression
framework based on pseudo-outcomes that use source outcomes to address
confounding and target covariates to define the population of interest.
We further propose a source-to-target extension of distance covariance
optimal weighting (DCOW), designed to remove treatment-covariate
dependence in the source sample while aligning the weighted source
covariate distribution with the target population. A central theoretical
contribution is a weight-level analysis of this optimization-based
procedure: we show that the population criterion identifies the oracle
source-to-target weights and that approximate empirical minimizers,
including exact minimizers as a special case, converge uniformly to
these weights under regularity conditions. We also establish consistency
and asymptotic normality of the resulting estimator. Simulations show
that the proposed method improves target dose-response estimation
relative to DCOW, generalized-propensity-score weighting, entropy
balancing, and unweighted alternatives. We illustrate the method in a
county-level analysis of PM\textsubscript{2.5} exposure and subsequent
heart-disease mortality using a source-target validation design.
\end{abstract}

\noindent {\it Keywords:} causal inference, average dose-response function, epigraphical convergence, non-convex optimization, two-sample U-processes, covariate shift
\vfill

\spacingset{1.6}
\section{Introduction}
\label{sec:intro}
Many scientific and policy questions concern the causal effect of a
continuous exposure or treatment. Examples include assessing the impact
of physician shortages on rural mortality rates \citep{gong2019higher},
studying the effect of air pollution on health outcomes
\citep{carre2017does}, and dose-response analyses in pharmaceutical
development \citep{holford1981understanding}. The usual estimand in this
setting is the average dose-response function (ADRF), which describes how
the mean potential outcome varies with the level of treatment. Early work
on multi-valued and continuous treatments introduced the generalized
propensity score \citep{imbens2000role,imai2004causal,
hirano2004propensity}. More recent work has developed nonparametric and
semiparametric estimators of the ADRF, including local smoothing
estimators based on pseudo-outcomes \citep{kennedy2017non}, kernel ridge
regression estimators \citep{singh2020kernel,kim2026estimating}, debiased and testing
procedures \citep{doss2024nonparametric,takatsu2022debiased}, and
related weighting and matching methods \citep{kallus2018confounding,
colangelo2020double,wu2024matching}.

Most existing ADRF methods assume that the observed sample represents the
population in which the dose-response function is to be estimated. This
assumption may fail in applications where outcome data are available from
a study or source population, while the population of scientific or policy
interest has a different covariate distribution. For example, evidence
from a clinical trial or observational study may need to be transported
to a regulatory population, a health system, or a geographic region with
different demographic and clinical characteristics
\citep{ramagopalan2022transportability,rubinstein2023balancing}. This
setting is closely related to causal generalization and transportability
\citep{degtiar2023review} and to covariate-shift problems in transfer
learning \citep{huang2006correcting,sugiyama2007direct,pan2009survey}.
We study the case in which covariates, treatment, and outcome are
observed in a source sample, but only covariates are observed in the
target sample. The goal is to use the labelled source data together with
the unlabelled target covariates to estimate the ADRF in the target
population. 

Covariate balancing by sample reweighting has been successful in causal
inference \citep{hainmueller2012entropy,imai2014covariate,
wang2020minimal,chen2023robust}. For continuous treatments without
covariate shift, \citet{huling2023independence} argue that weights should
make treatment and covariates independent in the weighted sample. Their
approach avoids explicit estimation of a high-dimensional conditional
treatment density, as in generalized-propensity-score methods, and does
not require the analyst to choose a finite set of covariate-treatment
moments to balance, as in some entropy-balancing approaches
\citep{fong2018covariate,tubbicke2021entropy,
vegetabile2021nonparametric}. Instead, treatment-covariate dependence is
measured through a distance-covariance criterion, which balances
distributions more globally.

Under covariate shift, however, independence weighting within the source
population is not sufficient. The weighted source sample must also
represent the target covariate distribution. Thus the weighting problem
has two simultaneous aims: to remove treatment-covariate dependence and
to transport the source covariate distribution to the target population.
Correcting only for source-population confounding can target the wrong
dose-response curve, and may even exacerbate bias under covariate shift
\citep{chen2023robust}. The central idea of this paper is to address
confounding adjustment and source-to-target transport in a single
weighting criterion, rather than first applying a source-population ADRF
method and then attempting a separate source-to-target adjustment.

We make three contributions. First, we formulate a new regression
framework, called \textit{Two-Sample Local Polynomial (TSLP) Regression},
for estimating the target-population ADRF from a labelled source sample
and an unlabelled target sample. TSLP extends the pseudo-outcome approach
of \citet{kennedy2017non} to the covariate-shifted setting. The
pseudo-outcome uses source outcomes and treatments to address confounding,
while the target covariates determine the population to which the
dose-response curve is transported. This construction leads to a
non-standard local polynomial estimator: the pseudo-outcomes share
information from the same target sample, so the estimator has the
structure of a generalized U-statistic formed from multiple independent
samples. We derive new asymptotic properties of TSLP estimators by
extending classical local polynomial theory to this two-sample setting
\citep{sen1974weak,lee2019u}.

Second, we propose a covariate-shifted extension of distance covariance
optimal weights (DCOW) \citep{huling2023independence} for constructing
the weighting component of the TSLP pseudo-outcome. The original DCOW
criterion was developed for continuous treatments in a single population
and targets treatment-covariate independence in the weighted sample. In
the source-to-target setting, this is not sufficient: the weighted source
sample must also represent the target covariate distribution. We therefore
augment the DCOW criterion with a source-to-target covariate balancing
component. The resulting weights are estimated directly from a
finite-sample optimization problem, without modelling the generalized
propensity score or inverting a high-dimensional conditional treatment
density. Empirically, TSLP with these weights performs well even with misspecified
outcome-regression augmentation. This behavior is
consistent with the doubly robust structure of the augmented
pseudo-outcome.

Third, we establish weight-level theoretical guarantees for the proposed
covariate-shifted DCOW criterion. At the population level, the criterion
identifies the oracle source-to-target weights needed to construct the
target-population pseudo-outcome. At the sample level, we show that
approximate empirical minimizers converge uniformly to these oracle
weights under regularity conditions. This result is distinct from the
existing DCOW theory, which studies properties of the resulting
dose-response estimator rather than establishing uniform convergence of
the estimated optimization weights themselves. Our proof uses epigraphical convergence
and provides a framework for analysing other optimization-based causal
weighting procedures. 

The numerical studies compare the proposed weights with DCOW,
generalized-propensity-score, entropy-balancing, and unweighted
alternatives under covariate shift. Across the sample sizes considered,
the proposed estimators have smaller bias and integrated estimation error
than the competing methods. We also apply the method to a county-level
analysis of PM\textsubscript{2.5} exposure and subsequent heart-disease
mortality, where the proposed transported estimates more closely track
an empirical target benchmark than the competing methods.

\section{Notation and problem setup}
\label{sec:problem_setup}

We describe the observed data structure and target parameter before
introducing the two-sample local polynomial estimator. Let
$(X,A,Y,S)$ be generated from a superpopulation $\mathbb P$, where
$X\in\mathcal X\subseteq\mathbb R^p$ denotes baseline covariates,
$A\in\mathcal A\subseteq\mathbb R$ denotes a continuous treatment,
$Y\in\mathbb R$ denotes the outcome, and $S\in\{1,0\}$ indicates source
or target membership. We call units with $S=1$ the source sample and
units with $S=0$ the target sample. The source and target sample sizes
are denoted by $n_S$ and $n_T$, respectively, with $n_S+n_T=N$. We
observe $(X,A,Y)$ for source units and only $X$ for target units. Thus
there are no treatment or outcome observations in the target sample. For
source units, we write $Z_i=(X_i,A_i,Y_i)$. When needed, target
covariates are denoted by $X_j^T$.

Let \(\mathbb P_S\) denote the conditional distribution of
\((X,A,Y)\) given \(S=1\). We write \(\mathbb P_S^X\) and
\(\mathbb P_T^X\) for the source and target covariate distributions,
respectively, and \(\mathbb P_S^A\) for the source treatment
distribution. The source joint distribution of \((X,A)\) is denoted by
\(\mathbb P_S^{X,A}\). The source and target covariate distributions are
not required to be equal.

When referring to densities, we use lowercase \(p\). For example,
\(p_S(x)\) and \(p_T(x)\) denote the source and target covariate
densities, \(p_S(a)\) denotes the source treatment density, and
\(p_S(x,a)\) denotes the source joint density of \((X,A)\). Since
treatment and outcome are observed only in the source population, any
observed marginal distribution involving \(A\) or \(Y\) is a source
marginal unless explicitly stated otherwise.

Empirical distributions are denoted by replacing \(\mathbb P\) with
\(\mathbb P_n\). For example, \(\mathbb P_{n_S}\) is the empirical
distribution of the labelled source sample, \(\mathbb P_{n_T}^X\) is the
empirical distribution of the target covariates, and
\(\mathbb P_{n_S}^A\) is the empirical distribution of the source
treatments. For a vector of source weights \(w\), we write
\(\mathbb P_{n_S}^{w}\) for the weighted empirical distribution of the
source \((X,A)\) sample.

We assume the following regularity conditions on the observed data. Let
$\tau^2(x,a)=\mathrm{var}(Y\mid X=x,A=a,S=1)$.

\begin{assumption}\label{finite_variance_assump}
The conditional variance satisfies $\tau^2(x,a)<\infty$.
\end{assumption}

\begin{assumption}\label{compact_A_assump}
The treatment space $\mathcal A$ is a compact subset of $\mathbb R$.
\end{assumption}

The compactness assumption restricts attention to a treatment region over
which the dose-response curve is to be estimated.

\subsection{Identification}
\label{sec:identification}

We use potential-outcome notation for continuous treatments
\citep{rubin2005causal}. Let $Y(a)$ be the potential outcome under
treatment value $a$, and let
\[
    \mu^*(x,a)=\mathbb E(Y\mid X=x,A=a,S=1)
\]
denote the source conditional outcome mean. The parameter of interest is
the target-population average dose-response function,
\[
    \theta(a_0)=\mathbb E_T\{Y(a_0)\}.
\]
The following assumptions identify this quantity from the source outcomes
and target covariates.

\begin{assumption}[Consistency]
\label{causal_assump_1}
$Y=Y(a)$ on the event $A=a$.
\end{assumption}

\begin{assumption}[Ignorability]
\label{causal_assump_2}
For all $a\in\mathcal A$, $Y(a)\perp A\mid X,S=1$.
\end{assumption}

\begin{assumption}[Transportability]
\label{causal_assump_3}
For all $a\in\mathcal A$ and covariate values in the target support,
\[
    \mathbb E\{Y(a)\mid X,S=0\}
    =
    \mathbb E\{Y(a)\mid X,S=1\}.
\]
\end{assumption}

The first two assumptions are standard causal identification assumptions
in the source population. The transportability assumption allows the
source and target covariate distributions to differ, but rules out
concept drift in the conditional mean potential outcome. Under
Assumptions~\ref{causal_assump_1}--\ref{causal_assump_3},
\[
    \theta(a_0)=\mathbb E_T\{\mu^*(X,a_0)\}.
\]

We also require overlap conditions for the weighting arguments below.

\begin{assumption}[Treatment overlap]
\label{causal_assump_4}
For all $a\in\mathcal A$ and all $x$ in the target covariate support,
the source conditional treatment density $p_S(a\mid x)$ is positive.
\end{assumption}

\begin{assumption}[Source participation overlap]
\label{causal_assump_5}
$p(S=1\mid X=x)>0$ for all $x$ with positive target density.
\end{assumption}

Because \(A\) is continuous, treatment overlap is a density condition.
In particular, the dose-response curve is interpreted over the treatment
region where the source conditional treatment density is positive for
target-relevant covariate values. The corresponding oracle
source-to-target weight is
\[
    w^*(x,a)
    =
    \frac{p_S(a)p_T(x)}{p_S(x,a)}
    =
    \frac{p_S(a)}{p_S(a\mid x)}
    \frac{p_T(x)}{p_S(x)} .
\]
Thus \(w^*\) is the usual stabilized inverse generalized propensity-score
weight in the source population, multiplied by the source-to-target
covariate density ratio.

\subsection{Two-sample local polynomial framework for covariate-shifted ADRF estimation}
\label{sec:tslp_framework}

We first recall the pseudo-outcome approach of \citet{kennedy2017non}
for estimating the ADRF in a single population. For a source observation
$(X_i,A_i,Y_i)$, the oracle pseudo-outcome is
\begin{align}
    (Y_i-\mu^*(X_i,A_i))
    \frac{p_S(A_i)p_S(X_i)}{p_S(X_i,A_i)}
    +
    \int \mu^*(x,A_i)\,d\mathbb P_S^X(x).
    \label{kennedy_pseudo_outcome}
\end{align}
In practice, the stabilized inverse generalized propensity-score weight
and the outcome regression are replaced by estimators $\hat w$ and
$\hat\mu$, and the integral is replaced by a source empirical average.
The resulting pseudo-outcome is then smoothed over the treatment values
using local polynomial regression.

In the covariate-shifted setting, the target ADRF requires the covariate
average in \eqref{kennedy_pseudo_outcome} to be taken over the target
population rather than the source population. Motivated by the oracle
weight $w^*$ above, we define the covariate-shifted sample
pseudo-outcome by
\begin{align}
    \hat\xi(Z_i,\hat w,\hat\mu)
    =
    \{Y_i-\hat\mu(X_i,A_i)\}\hat w(X_i,A_i)
    +
    \frac{1}{n_T}\sum_{j=1}^{n_T}\hat\mu(X_j^T,A_i).
    \label{cheng_pseudo_outcome}
\end{align}
The first term uses the labelled source data to adjust for confounding;
the second term uses the target covariates to define the population to
which the dose-response curve is transported. This pseudo-outcome has the
same double-robust structure as the single-population pseudo-outcome. To
state this property without treating the estimated nuisance functions as
fixed, let
\[
    \xi_{w,\mu}(Z)
    =
    \{Y-\mu(X,A)\}w(X,A)
    +
    \int \mu(x,A)\,d\mathbb P_T^X(x)
\]
be the population analogue formed with fixed nuisance functions \(w\) and
\(\mu\). For a fixed treatment value $a_0$,
\[
    \mathbb E\{\xi_{w,\mu}(Z)\mid A=a_0\}=\theta(a_0)
\]
if either \(w=w^*\) or \(\mu=\mu^*\). The sample pseudo-outcome in
\eqref{cheng_pseudo_outcome} replaces the target integral by the target
empirical average and replaces the nuisance functions by estimates. A
formal proof is given in Supplementary
Section~A.1. The efficient
influence-function calculation for the corresponding integrated target
ADRF is given in Supplementary Section~A.2.

Given the pseudo-outcomes, the local linear TSLP estimator at treatment
level $a_0$ is
\begin{align}
    (\hat\theta(a_0),\hat\theta'(a_0))^\top
    =
    \arg\min_{a,b\in\mathbb R}
    \mathbb P_{n_S}
    \left[
        K_{ha_0}(A)
        \left\{
            \hat\xi(Z,\hat w,\hat\mu)
            -a-b(A-a_0)/h
        \right\}^2
    \right].
    \label{local_linear_regression_objective}
\end{align}
Here $K_{ha_0}(A)$ is the kernel weight centred at $a_0$. Higher-order
local polynomial versions are defined analogously.

Although \eqref{local_linear_regression_objective} has the form of a
local polynomial regression, it is not a standard one-sample smoother.
Each pseudo-outcome in \eqref{cheng_pseudo_outcome} contains an empirical
average over the target covariates, inducing shared information across the pseudo-outcomes. Drawing on the structure of two-sample U-statistics, we refer to the resulting estimator as a Two-Sample Local Polynomial (TSLP) estimator. Because these correlations preclude standard one-sample asymptotic theory, we develop a novel two-sample argument in Section~\ref{sec:asymptotics}.

\section{Properties of a suitable weighting method}
\label{sec:weighting_criterion}

\subsection{Why the target product distribution matters}
\label{sec:bias_weighting_target}

The pseudo-outcome in \eqref{cheng_pseudo_outcome} combines a weighted
residual term from the labelled source sample with an average of the
outcome regression over the target covariates. This structure determines
the distributional target of the weights. Let \(\theta^*(a_0)\) be the
idealized TSLP estimator formed from pseudo-outcomes that use the oracle
weight \(w^*\) and the true outcome regression \(\mu^*\). Then
\begin{align*}
    \hat{\theta}(a_{0}) - \theta(a_{0}) = \underbrace{\theta^{*}(a_{0}) - \theta(a_{0})}_{\text{oracle error}} + \underbrace{\hat{\theta}(a_{0}) - \theta^{*}(a_{0})}_{\text{nuisance error}}.
\end{align*}
The first term is the oracle smoothing error. The second term is the
nuisance error caused by estimating the weight and outcome-regression
components.

Up to the standard bounded local-polynomial normalising factor, the
leading nuisance term has the form
\[
    \mathbb P_{n_S}
    \left[
        g_{ha_0}(A)K_{ha_0}(A)
        \{\hat\xi(Z,\hat w,\hat\mu)-\hat\xi(Z,w^*,\mu^*)\}
    \right],
\]
where \(g_{ha_0}\), \(K_{ha_0}\), and the normalising matrix are the usual
local-polynomial quantities, and \(\hat\xi(Z,w,\mu)\) denotes the
pseudo-outcome in \eqref{cheng_pseudo_outcome} evaluated with generic
nuisance functions \(w\) and \(\mu\). A direct expansion, given in
Supplementary Section~B.1, shows that
this term contains
\[
\begin{aligned}
&\frac{1}{n_S}\sum_{i=1}^{n_S}
    g_{ha_0}(A_i)K_{ha_0}(A_i)
    \varepsilon_i\{\hat w(X_i,A_i)-w^*(X_i,A_i)\}  \\
&\quad+
\int g_{ha_0}(a)K_{ha_0}(a)
    \{\mu^*(x,a)-\hat\mu(x,a)\}
    \left[
        d\mathbb P_{n_S}^{\hat w}(x,a)
        -d\mathbb P_{n_T}^X(x)d\mathbb P_{n_S}^A(a)
    \right],
\end{aligned}
\]
where \(\varepsilon_i=Y_i-\mu^*(X_i,A_i)\). The first term is a
residual-weight interaction. The second term shows that, when the outcome
regression is imperfect, the relevant discrepancy is between the weighted
source empirical distribution of covariates and treatment and the product
of the target empirical covariate distribution and the source empirical
treatment distribution.

Thus a suitable weighting method should not only reduce
treatment-covariate dependence in the source sample. It should also
transport the source covariate distribution to the target population. The
criterion below measures these two requirements by combining a
covariate-shifted distance-covariance component with marginal matching
terms based on energy distance.

\subsection{Distance-covariance component}
\label{sec:shifted_dcov}

Distance covariance measures dependence by comparing a joint
characteristic function with the product of its marginal characteristic
functions \citep{szekely2007measuring}. DCOW
\citep{huling2023independence} uses this idea to choose weights that
make treatment and covariates approximately independent in a single
population. In the present source-to-target problem, the desired product
distribution has the target covariate marginal and the source treatment
marginal.

For a nonnegative source weight vector \(w=(w_1,\ldots,w_{n_S})^\top\)
with \(\sum_i w_i=n_S\), let \(\phi^{n_S}_{X,A,w}(t,s)\) denote the
weighted empirical characteristic function of the source \((X,A)\)
sample, with analogous notation for \(\phi^{n_S}_{X,w}(t)\) and
\(\phi^{n_S}_{A,w}(s)\). Let \(\phi^{n_T}_{X_T}(t)\) be the empirical
characteristic function of the target covariates, and let
\(\phi^{n_S}_{A}(s)\) be the unweighted empirical characteristic function
of the source treatment. The notation is summarized in Supplementary
Section~B.2. Define
\[
\begin{aligned}
    \mathcal V_N^2(X,A,X_T,w)
    ={}&
    \int_{\mathbb R^{p+1}}
    \Big|
        \{\phi^{n_S}_{X,A,w}(t,s)
          -\phi^{n_T}_{X_T}(t)\phi^{n_S}_{A,w}(s)\}
        +\{\phi^{n_T}_{X_T}(t)-\phi^{n_S}_{X,w}(t)\}
          \phi^{n_S}_{A}(s)
    \Big|^2
    \omega(t,s)\,dt\,ds,
\end{aligned}
\]
where
\[
    \omega(t,s)=\{c_pc_1\|t\|_2^{p+1}|s|^2\}^{-1},
    \qquad
    c_d=\frac{\pi^{(1+d)/2}}{\Gamma\{(1+d)/2\}}.
\]
The first part of the integrand measures departure from independence in
the weighted source distribution. The second part shifts the covariate
margin in this comparison from the source sample to the target sample.
If the target covariate sample is replaced by the source covariate
sample, the criterion reduces to the single-population DCOW criterion, up
to notational conventions.

\subsection{Marginal matching component}
\label{sec:energy_matching}

The distance-covariance component alone does not ensure that the
weighted marginals equal the desired marginals. We therefore add two
weighted energy-distance terms \citep{szekely2013energy,huling2020energy}:
\[
    \mathcal E_N(X_w,X_T)
    =
    \int_{\mathbb R^p}
    \left|\phi^{n_S}_{X,w}(t)-\phi^{n_T}_{X_T}(t)\right|^2
    \{c_p\|t\|_2^{p+1}\}^{-1}\,dt,
\]
\[
    \mathcal E_N(A_w,A)
    =
    \int_{\mathbb R}
    \left|\phi^{n_S}_{A,w}(s)-\phi^{n_S}_{A}(s)\right|^2
    \{c_1|s|^2\}^{-1}\,ds.
\]
The first term matches the weighted source covariate marginal to the
target covariate marginal. The second preserves the source treatment
marginal.

\subsection{Combined criterion and computational form}
\label{sec:combined_criterion}

Combining the dependence and marginal-matching components gives the
finite-sample criterion.
\begin{definition}[$D_N(w)$]
\label{def:DN_w}
Given source covariates \(X\), source treatments \(A\), target covariates
\(X_T\), and a nonnegative weight vector \(w\) satisfying
\(\sum_{i=1}^{n_S}w_i=n_S\), define
\[
    D_N(w)
    =
    \mathcal V_N^2(X,A,X_T,w)
    +\mathcal E_N(X_w,X_T)
    +\mathcal E_N(A_w,A).
\]
\end{definition}

The following proposition characterizes the finite-sample target of the
criterion. Its proof is given in Supplementary
Section~B.3.
\begin{proposition}[Target-product characterisation]
\label{prop:bound_on_cf_diff}
Let \(w\) be a nonnegative weight vector satisfying
\(\sum_{i=1}^{n_S}w_i=n_S\). Then \(D_N(w)\geq0\), and \(D_N(w)=0\) if
and only if
\[
    \phi^{n_S}_{X,A,w}(t,s)
    =
    \phi^{n_T}_{X_T}(t)\phi^{n_S}_{A}(s)
    \qquad
    \text{for all }(t,s)\in\mathbb R^{p+1}.
\]
\end{proposition}

Proposition~\ref{prop:bound_on_cf_diff} shows that the zero of
\(D_N(w)\) corresponds to the desired product distribution: target
covariates and source treatments, with no treatment-covariate dependence
after reweighting. Minimizing \(D_N(w)\) is therefore a finite-sample
analogue of the oracle weighting target identified in
Section~\ref{sec:problem_setup}.

Although \(D_N(w)\) is defined through empirical characteristic
functions, it has an equivalent pairwise Euclidean-distance
representation. The next two propositions summarize the computational
forms; explicit expressions are given in Supplementary
Section~B.4.
\begin{proposition}[Pairwise-distance form of \(\mathcal V_N^2\)]
\label{prop:euclid_diff_V2}
For fixed source and target samples, \(\mathcal V_N^2(X,A,X_T,w)\) can
be written as a quadratic function of the weight vector \(w\), with
coefficients determined by pairwise distances among source covariates,
target covariates, and source treatment values.
\end{proposition}

\begin{proposition}[Pairwise-distance form of the energy distances]
\label{prop:euclid_diff_energy}
For fixed source and target samples, \(\mathcal E_N(X_w,X_T)\) and
\(\mathcal E_N(A_w,A)\) can be written as quadratic functions of the
weight vector \(w\), with coefficients determined by the corresponding
source-source, source-target, and target-target pairwise distances.
\end{proposition}

Consequently, for fixed source and target samples, there exist a
symmetric matrix \(Q_N\), a vector \(q_N\), and a scalar \(c_N\), all
computed from pairwise distances, such that
\[
    D_N(w)=w^\top Q_Nw+q_N^\top w+c_N.
\]
This quadratic representation is the basis for the finite-sample
optimization problem in Section~\ref{sec:opt_weights}. The matrix
\(Q_N\) need not be positive semidefinite, so the resulting quadratic
program is not necessarily convex.

\section{Optimization weights}
\label{sec:opt_weights}

\subsection{Finite-sample optimization problem}
\label{section:optimization_problem}

The criterion \(D_N(w)\) defined in Section~\ref{sec:combined_criterion}
measures the discrepancy between the weighted source distribution of
\((X,A)\) and the product of the target covariate distribution and the
source treatment distribution. We estimate the weights by solving
\begin{subequations}
\label{weights_opt_problem}
\begin{alignat}{2}
&\min_{w\in\mathbb R^{n_S}}        &\qquad& D_N(w)  \\
&\text{subject to}                 &      & \sum_{i=1}^{n_S} w_i=n_S, \\
&                                  &      & 0\leq w_i\leq M,\qquad i=1,\ldots,n_S .
\end{alignat}
\end{subequations}
The normalization makes the weighted empirical source distribution a
probability distribution. The upper bound \(M\) prevents individual
observations from receiving arbitrarily large weight. This constraint is
also consistent with the regularity condition that the oracle
source-to-target density ratio \(w^*\) is uniformly bounded. In the
numerical studies, we use a loose upper bound,
\(M=\max(500,n_S/4)\), and the fitted weights rarely approach this
limit.

By the pairwise-distance representation of
Section~\ref{sec:combined_criterion}, \(D_N(w)\) is a quadratic function
of the weight vector for fixed source and target samples. The resulting
optimization problem is therefore a quadratic program with linear
constraints, although the quadratic matrix need not be positive
semidefinite. In our implementation, we use OSQP
\citep{stellato2020osqp} to obtain a computationally efficient
approximate solution. Although OSQP is designed for convex quadratic
programs whereas the present objective is not necessarily convex, it
performed stably in our numerical experiments and offered substantial
computational advantages. The theory below is formulated for approximate
minimizers, with exact minimizers included as the special case
\(\epsilon_N=0\). When the numerical solution satisfies the
approximate-minimizer condition with tolerance \(\epsilon_N\) and
\(\epsilon_N\to0\), the resulting sequence of weights converges uniformly
to the oracle source-to-target weights under the stated regularity
conditions.

\subsection{From sample weights to a weighting function}
\label{sec:sample_weights_to_function}

The finite optimization problem returns a vector of weights at the
observed source sample points. To study uniform convergence to the oracle
weight, we represent such a vector as the point evaluations of a weighting
function on \(\Omega=\mathcal X\times\mathcal A\). Specifically, a
candidate weighting function \(w\) induces the sample weight vector
\[
    \{w(X_1,A_1),\ldots,w(X_{n_S},A_{n_S})\}^\top,
\]
and, with a slight abuse of notation, we write
\[
    D_N(w)
    =
    D_N\bigl(w(X_1,A_1),\ldots,w(X_{n_S},A_{n_S})\bigr).
\]
Conversely, any feasible vector satisfying
\(\sum_i w_i=n_S\), can be represented by a smooth interpolating function
\(\widetilde w_N\) such that
\[
    \widetilde w_N(X_i,A_i)=w_i,\qquad i=1,\ldots,n_S,
\]
and
\[
    \int_{\Omega}\widetilde w_N(x,a)\,d\mathbb P_S^{X,A}(x,a)=1.
\]
The construction uses disjoint smooth bump functions centred at the
observed source points and is given in Supplementary
Section~C.1. This interpolation establishes a
formal pointwise correspondence between a finite feasible vector and a
smooth function. The convergence theorem below is therefore stated for approximate
minimizers in the admissible Sobolev class, while the numerical estimator uses
the finite weight vector from \eqref{weights_opt_problem} directly.

Let \(\mathcal W\) denote the class of admissible weighting functions on
\(\Omega\). In the theory below, functions in \(\mathcal W\) are
nonnegative, bounded by \(M\), normalized to integrate to one with
respect to \(\mathbb P_S^{X,A}\), and belong to a Sobolev class with
sufficient smoothness for uniform convergence. This function-space
formulation allows us to state convergence of the estimated weights to
the oracle source-to-target weight
\[
    w^*(x,a)=\frac{p_T(x)p_S(a)}{p_S(x,a)}.
\]

\subsection{Population target and uniform convergence}
\label{sec:weight_convergence}

The population analogue of \(D_N(w)\) is
\[
\begin{aligned}
D(w)
={}&
\mathcal V^2(F_{X,A,w},F_{X_T},F_A)
+\mathcal E(F_{X,w},F_{X_T})
+\mathcal E(F_{A,w},F_A)  \\
={}&
\int_{\mathbb R^{p+1}}
\left|
\phi_{X,A,w}(t,s)
-\phi_{X_T}(t)\phi_{A,w}(s)
+\{\phi_{X_T}(t)-\phi_{X,w}(t)\}\phi_A(s)
\right|^2
\omega(t,s)\,dt\,ds  \\
&+
\int_{\mathbb R^p}
\left|\phi_{X,w}(t)-\phi_{X_T}(t)\right|^2\omega_X(t)\,dt
+
\int_{\mathbb R}
\left|\phi_{A,w}(s)-\phi_A(s)\right|^2\omega_A(s)\,ds .
\end{aligned}
\]
Here \(F_A\) and \(\phi_A\) refer to the source treatment marginal,
since treatment is not observed in the target sample. The first term
measures departure from the desired product structure, and the two
energy-distance terms enforce the target covariate marginal and the
source treatment marginal.

\begin{lemma}[Population minimizer]
\label{lem:unique_min}
The population criterion \(D(w)\) has a unique minimizer over
\(\mathcal W\), up to \(\mathbb P_S^{X,A}\)-null sets, and this minimizer is
\[
    w^*(x,a)=\frac{p_T(x)p_S(a)}{p_S(x,a)} .
\]
\end{lemma}

The proof is given in Supplementary Section~C.5. The
lemma shows that the population version of the proposed criterion targets
the source-to-target reweighting needed for the target ADRF.

We next state the convergence result. Let \(q>p+1\), where \(p\) is the
dimension of \(X\), and view the weights as functions on
\(\Omega=\mathcal X\times\mathcal A\).

\begin{assumption}[Sobolev regularity of the oracle weight]
\label{wstar_sobolev}
The oracle weight \(w^*\) belongs to \(\mathcal W^{1,q}(\Omega)\) for
some \(q>p+1\).
\end{assumption}

\begin{assumption}[Domain regularity]
\label{Omega_boundary}
The domain \(\Omega=\mathcal X\times\mathcal A\) is bounded and has a
Lipschitz boundary.
\end{assumption}

The Lipschitz-boundary condition is used to apply standard Sobolev
embedding results. A formal definition is given in Supplementary
Section~C.3.

\begin{theorem}[Uniform convergence of optimization weights]
\label{thm:weight_unif_conv}
Let \(\epsilon_N\) be a sequence of nonnegative random variables with
\(\epsilon_N\to0\) almost surely. Let \(w_N\in\mathcal W\) be a sequence
of \(\epsilon_N\)-tolerant minimizers of the empirical criterion, so that
\[
    D_N(w_N)
    \leq
    \inf_{w\in\mathcal W}D_N(w)+\epsilon_N .
\]
Under Assumptions~\ref{wstar_sobolev} and~\ref{Omega_boundary}, with
probability one,
\[
    \|w_N-w^*\|_\infty\to0 .
\]
\end{theorem}

The proof is given in Supplementary Section~C.6.
It proceeds by showing epigraphical convergence of the empirical
criterion to the population criterion, identifying the unique population
minimizer by Lemma~\ref{lem:unique_min}, and then using Sobolev embedding
to strengthen subsequential convergence to uniform convergence. The
result is a weight-level guarantee: it studies the estimated optimization
weights themselves, rather than only the final smoothed ADRF estimator.

As a consequence, the weighted source empirical distribution converges to
the product distribution that appears in the target ADRF pseudo-outcome.

\begin{corollary}[Weighted source distribution]
\label{cor:convergence_in_distribution}
Let \(w_N\) satisfy the conditions of
Theorem~\ref{thm:weight_unif_conv}. Then, with probability one,
\[
    \mathbb P_{n_S}^{w_N}
    \rightsquigarrow
    \mathbb P_T^X\otimes\mathbb P_S^A.
\]
\end{corollary}

The proof is given in Supplementary
Section~C.8. This corollary connects the
weight-level convergence theorem to the product-distribution target used
in the TSLP pseudo-outcome. In Section~\ref{sec:asymptotics}, we combine
this result with the two-sample local-polynomial analysis to establish
the large-sample properties of the proposed ADRF estimator.

\section{Asymptotic properties of the TSLP estimator}\label{sec:asymptotics}

Section~\ref{sec:opt_weights} establishes weight-level convergence for the
proposed source-to-target weighting criterion. We now study the large-sample
properties of the two-sample local polynomial (TSLP) estimator. The results in
this section are stated for generic nuisance estimators \(\hat w\) and
\(\hat\mu\). The proposed weights enter through the high-level condition that
\(\hat w\) converges to the oracle source-to-target weight \(w^*\).

Let \(\bar w\) and \(\bar\mu\) denote probability limits of \(\hat w\) and
\(\hat\mu\). Define the limiting pseudo-outcome by
\[
    \hat\xi(Z_i,\bar w,\bar\mu)
    =
    \{Y_i-\bar\mu(X_i,A_i)\}\bar w(X_i,A_i)
    +
    \frac{1}{n_T}\sum_{j=1}^{n_T}\bar\mu(X_j^T,A_i).
\]
The corresponding infeasible TSLP estimator, hereafter referred to as the limiting estimator, is
\[
    \tilde\theta(a_0)
    =
    g_{ha_0}(a_0)^\top \hat D_{ha_0}^{-1}
    \mathbb P_{n_S}
    \left[
        g_{ha_0}(A)K_{ha_0}(A)
        \hat\xi(Z,\bar w,\bar\mu)
    \right],
\]
where \(g_{ha_0}(A)=(1,(A-a_0)/h)^\top\),
\(K_{ha_0}(A)=h^{-1}K\{(A-a_0)/h\}\), and \(\hat D_{ha_0}\) is the usual
local-polynomial design matrix. We decompose
\[
    \hat\theta(a_0)-\theta(a_0)
    =
    \{\tilde\theta(a_0)-\theta(a_0)\}
    +
    \{\hat\theta(a_0)-\tilde\theta(a_0)\}.
\]
The first term is the limiting two-sample local-polynomial error, and the
second term is the nuisance discrepancy. Section~\ref{sec:bias_weighting_target}
uses a related decomposition to motivate the weighting target; here the
purpose is different, namely to separate the limiting distribution of the
infeasible smoother from the error caused by estimating the nuisance functions.

Unlike a standard one-sample local polynomial estimator, \(\tilde\theta(a_0)\)
uses pseudo-outcomes that share the same target covariates. The leading term is
therefore a generalized U-statistic formed from one source sample and one target
sample. This two-sample structure is responsible for the asymptotic results
below.

\begin{assumption}\label{smoothness_assumptions}
The target ADRF \(\theta(a)\) is twice continuously differentiable in a
neighbourhood of \(a_0\), and the source treatment density \(p_S(a)\) is
continuously differentiable with \(p_S(a_0)>0\).
\end{assumption}

\begin{assumption}\label{bandwidth_assumptions}
The bandwidth satisfies \(h\to0\) and \(n_S h\to\infty\). In addition,
\(n_S/(n_S+n_T)\to\kappa\), where \(0<\kappa<1\).
\end{assumption}

\begin{assumption}\label{kernel_assumptions}
The kernel \(K\) is symmetric and satisfies \(\int K(u)du=1\),
\(s_2=\int u^2K(u)du<\infty\), \(\int K(u)^2du<\infty\), and
\(\int u^4K(u)^2du<\infty\).
\end{assumption}

\begin{proposition}[Asymptotic normality of the limiting estimator]
\label{prop:asymptotic_normality_limiting_estimator}
Let \(\bar w\) and \(\bar\mu\) be fixed functions such that either
\(\bar w=w^*\) or \(\bar\mu=\mu^*\). Under
Assumptions~\ref{smoothness_assumptions}--\ref{kernel_assumptions},
\[
    \sqrt{n_S h}
    \left\{
        \tilde\theta(a_0)-\theta(a_0)
        -\frac{h^2}{2}\theta''(a_0)s_2
    \right\}
    \rightsquigarrow
    N\{0,\delta_{1,0}^2(a_0)\},
\]
where
\[
    \delta_{1,0}^2(a_0)
    =
    \{\gamma(a_0)+\sigma(a_0)\}
    \frac{\int K(u)^2du}{p_S(a_0)},
\]
\[
    \sigma(a_0)
    =
    \mathbb E\{\tau^2(X,A)\bar w(X,A)^2\mid A=a_0\},
    \qquad
    \tau^2(X,A)=\operatorname{var}(Y\mid X,A,S=1),
\]
and
\[
    \gamma(a_0)
    =
    \operatorname{var}
    \left[
        \{\mu^*(X,A)-\bar\mu(X,A)\}\bar w(X,A)
        \mid A=a_0
    \right].
\]
\end{proposition}

The proof of Proposition~\ref{prop:asymptotic_normality_limiting_estimator}
uses the central limit theorem for generalized U-statistics of two independent
samples. The target-sample projection is asymptotically degenerate after local
smoothing, so the leading variance depends on the source sample projection.
This is why the normalization is \(\sqrt{n_S h}\).

We use standard empirical-process notation for the nuisance-function conditions.
An envelope for a function class \(\mathcal F\) is a function \(F\) such that
\(|f|\le F\) for all \(f\in\mathcal F\). For a probability measure \(Q\), let
\(N_p(\varepsilon,Q,\mathcal F,F)\) denote the corresponding \(L_p(Q)\)
covering number normalized by the envelope. Define
\[
    \mathcal J_{m,p}(\gamma,\mathcal F,F)
    =
    \int_0^\gamma
    \sup_Q
    \{1+\log N_p(\varepsilon,Q,\mathcal F,F)\}^{m/2}
    \,d\varepsilon .
\]
A function class is called VC if its covering numbers are polynomially bounded
in \(1/\varepsilon\).

\begin{assumption}\label{uniform_bounded_assump}
The source joint density \(p_S(x,a)\), the oracle weight \(w^*(x,a)\), and the
outcome regression \(\mu^*(x,a)\) are uniformly bounded on their supports.
\end{assumption}

\begin{assumption}\label{kernel_assump_2}
The kernel \(K\) is continuous, symmetric, and compactly supported on \([-1,1]\).
The kernel classes generated by \(K_{ha_0}(A)\) and
\((A-a_0)K_{ha_0}(A)\), indexed by \(a_0\in\mathcal A\), are VC classes.
\end{assumption}

\begin{assumption}\label{entropy_integral_assump}
The nuisance estimators and their limits belong to uniformly bounded function
classes: \(\hat w,\bar w\in\mathcal F_w\) with envelope \(F_w\), and
\(\hat\mu,\bar\mu\in\mathcal F_\mu\) with envelope \(F_\mu\). The relevant
entropy integrals are finite; in particular,
\(\mathcal J_{2,2}(1,\mathcal F_w,F_w)<\infty\) and
\(\mathcal J_{2,2}(1,\mathcal F_\mu,F_\mu)<\infty\).
\end{assumption}

For local rates, let \(\|\cdot\|_{2,a}\) denote the
\(L_2(\mathbb P_S^{X\mid A=a})\)
norm. Suppose that, for \(a\) in a neighbourhood of \(a_0\),
\[
    \sup_{|t-a_0|\le h}\|\hat w(\cdot,t)-w^*(\cdot,t)\|_{2,t}
    =O_p\{r_N(a_0)\},
\]
and
\[
    \sup_{|t-a_0|\le h}\|\hat\mu(\cdot,t)-\mu^*(\cdot,t)\|_{2,t}
    =O_p\{s_N(a_0)\}.
\]

\begin{theorem}[Consistency rate of the TSLP estimator]
\label{thm:consistency_rate}
Assume the conditions of Proposition~\ref{prop:asymptotic_normality_limiting_estimator}
and Assumptions~\ref{uniform_bounded_assump}--\ref{entropy_integral_assump}. If
\(\hat w\to\bar w\) and \(\hat\mu\to\bar\mu\) in the nuisance-function classes
and either \(\bar w=w^*\) or \(\bar\mu=\mu^*\), then
\[
    |\hat\theta(a_0)-\theta(a_0)|
    =
    O_p\left(
        \frac{1}{\sqrt{n_S h}}
        +
        \frac{1}{\sqrt{n_T}}
        +
        h^2
        +
        r_N(a_0)s_N(a_0)
    \right).
\]
\end{theorem}

The term \(1/\sqrt{n_S h}\) is the usual local-smoothing stochastic order from
the labelled source sample. The term \(1/\sqrt{n_T}\) reflects the empirical
target-covariate average in the pseudo-outcome. Under
Assumption~\ref{bandwidth_assumptions}, the source local-smoothing term is the
leading stochastic term in the displayed rate. The product
\(r_N(a_0)s_N(a_0)\) is the usual second-order nuisance term from the augmented
pseudo-outcome. Thus consistency can be obtained either from a consistent
weighting component with a bounded outcome-regression limit, or from a
consistent outcome regression with a bounded weighting limit; when both
components improve, the product term becomes smaller.

For asymptotic normality, the nuisance discrepancy must be negligible on the
\(\sqrt{n_S h}\) scale. We state this as a high-level local
stochastic-equicontinuity condition, together with the second-order
product-rate requirement.

\begin{assumption}[Negligible nuisance discrepancy]
\label{nuisance_asymp_negligible}
At the treatment value \(a_0\),
\[
    \sqrt{n_S h}\{\hat\theta(a_0)-\tilde\theta(a_0)\}=o_p(1).
\]
A sufficient condition is that the empirical-process part of the nuisance
discrepancy is \(o_p\{(n_Sh)^{-1/2}\}\) and
\(r_N(a_0)s_N(a_0)=o_p\{(n_Sh)^{-1/2}\}\).
\end{assumption}

\begin{theorem}[Asymptotic normality of the TSLP estimator]
\label{thm:asymptotic_normality}
Assume the conditions of Proposition~\ref{prop:asymptotic_normality_limiting_estimator}
and Assumption~\ref{nuisance_asymp_negligible}. Then
\[
    \sqrt{n_S h}
    \left\{
        \hat\theta(a_0)-\theta(a_0)
        -\frac{h^2}{2}\theta''(a_0)s_2
    \right\}
    \rightsquigarrow
    N\{0,\delta_{1,0}^2(a_0)\},
\]
with \(\delta_{1,0}^2(a_0)\) as defined in
Proposition~\ref{prop:asymptotic_normality_limiting_estimator}.
\end{theorem}

\section{Simulation study}
\label{sec:simulation}

We use a synthetic experiment to evaluate the effect of the weighting method,
outcome augmentation, and source-to-target adaptation on estimation of the target
ADRF. The data-generating mechanism follows \citet{vegetabile2021nonparametric},
with a modification that introduces covariate shift between the labelled source
sample and the unlabelled target sample. The proposed estimator is compared with
generalized-propensity-score methods, entropy-balancing methods, their
source-to-target adapted versions, DCOW, and an unweighted baseline.

\subsection{Data generation}
\label{sec:simulation-dgp}

The source latent covariates are generated as
\[
X_1\sim N(-0.5,1),\quad X_2\sim N(1,1),\quad
X_3\sim N(0,1),\quad X_4\sim N(1,1),\quad
X_5\sim \mathrm{Bernoulli}(0.3).
\]
The target latent covariates are generated from shifted marginal distributions,
\[
X_1\sim N(0,1),\quad X_2\sim N(1.5,1),\quad
X_3\sim N(0.5,1),\quad X_4\sim N(1.5,1),\quad
X_5\sim \mathrm{Bernoulli}(0.4).
\]
For both samples, the covariates supplied to the estimators are nonlinear
transformations of these latent variables,
\begin{align*}
V_1&=\exp(X_1/2),
&
V_2&=\frac{X_2}{1+\exp(X_1)}+10,\\
V_3&=\frac{X_1X_3}{25}+0.6,
&
V_4&=\{X_4-E(X_4\mid S)\}^2,
&
V_5&=X_5 .
\end{align*}
Here the conditional mean in the definition of \(V_4\) is taken under the
corresponding source or target covariate distribution. 

Treatment is generated from a noncentral chi-square distribution with three
degrees of freedom and noncentrality parameter
\[
    \mu_A(X)=5|X_1|+6|X_2|+|X_4|+3|X_5|.
\]
The outcome is generated from
\[
Y=\frac{1}{50}\left[-0.15A^2
    +A(X_1^2+X_2^2)
    +(X_1+3)^2+2(X_2-25)^2+X_3-C+\varepsilon\right],
\quad \varepsilon\sim N(0,1),
\]
where
\[
    C=(-0.5+3)^2+2(1-25)^2+18.
\]
The transformations induce nonlinear relationships among the observed
covariates, treatment, and outcome. The shift in the covariate distribution is
chosen to create a nontrivial source-to-target transport problem while retaining
adequate overlap over the treatment range considered.

We vary the labelled source sample size over
\(n_S\in\{250,500,1000,2000\}\) and the unlabelled target sample size over
\(n_T\in\{250,500,1000\}\). As in \citet{vegetabile2021nonparametric}, we
estimate the ADRF on the treatment interval \([1.5,45]\). The target ADRF used
for evaluation is computed by Monte Carlo using a large independent sample from
the target covariate distribution.

\subsection{Comparison methods}
\label{sec:sim_comparison_methods}

The comparison methods are chosen to represent four relevant strategies:
distance-covariance independence weighting, generalized-propensity-score
weighting for continuous treatments, moment-balancing weights, and
two-step adaptations that combine a source-population continuous-treatment
method with a separate source-to-target covariate adjustment. This allows
us to assess whether the proposed one-step criterion improves over methods
that address only treatment-covariate dependence or handle covariate shift
through a separate adaptation step.

We first compare with the original distance covariance optimal weights
(DCOW) of \citet{huling2023independence}. DCOW is the most direct
comparison of the proposed method: it targets treatment-covariate
independence in the source sample using a distance-covariance criterion,
but it does not include the source-to-target covariate balancing
component. Thus the comparison with DCOW isolates the effect of adding
target-population covariate balance to the independence-weighting
criterion.

We next compare with generalized-propensity-score methods. GPS-LM
estimates the conditional treatment density using a linear regression of
treatment on covariates and a Gaussian residual model. GPS-BART uses a
Bayesian additive regression tree model for the conditional mean of the
treatment and the same residual-density construction. Their adapted
versions multiply the stabilized generalized propensity-score weights by
an estimated source-to-target covariate density ratio. In our
implementation, this density ratio is estimated by a logistic-regression
classifier distinguishing target units from source units; further details
are given in Supplementary
Section~E.1.

We also compare with entropy balancing for continuous treatments
\citep{tubbicke2021entropy,vegetabile2021nonparametric}. This method
chooses nonnegative weights by minimizing relative entropy subject to
moment-balance constraints involving powers and interactions of centred
treatment and covariate terms. In our implementation, the moment order is
set to two. The adapted entropy method first aligns source covariates
with target covariates by binary entropy balancing and then applies
continuous-treatment entropy balancing using the first-step weights as
base weights. This comparison evaluates whether explicit moment balancing
can recover the target ADRF under covariate shift. The unweighted
estimator is included to show the effect of ignoring both confounding and
covariate shift.

For each weighting method, we report both an unaugmented estimator and an
outcome-augmented estimator. The augmented estimator uses an ordinary
least squares outcome regression in the pseudo-outcome. For the
comparison methods, we use a residual-smoothing form: local linear
regression is applied to the weighted residual component, and the
target-sample average of the fitted outcome regression is added after
smoothing. The local polynomial fit is local linear throughout.

We evaluate performance using mean absolute bias (MAB) and integrated root mean
squared error (IRMSE),
\begin{align*}
    \mathrm{MAB}
    &= \int_{\mathcal A}
    \left|\frac{1}{R}\sum_{r=1}^R\widehat\theta_r(a)-\theta(a)\right|
    \widehat p_S(a)\,da,\\
    \mathrm{IRMSE}
    &= \int_{\mathcal A}
    \left\{\frac{1}{R}\sum_{r=1}^R
    \bigl(\widehat\theta_r(a)-\theta(a)\bigr)^2\right\}^{1/2}
    \widehat p_S(a)\,da .
\end{align*}
Here \(R=500\) is the number of Monte Carlo replications,
\(\widehat p_S(a)\) is an estimate of the source treatment density, and
\(\theta(a)\) denotes the target ADRF. The weighting by \(\widehat p_S\) focuses
the summary metrics on the treatment region well represented in the source
sample.

\subsection{Simulation results}\label{sec:simulation_results}

Tables~\ref{tab:simulation-mab-main} and
\ref{tab:simulation-irmse-main} report the Monte Carlo performance,
averaged over \(n_T\in\{250,500,1000\}\) within each source sample size.
The full results stratified by both \(n_S\) and \(n_T\) are reported in
Supplementary Tables~S.1--S.4.
The tables compare three aspects of performance: the effect of adding
target covariate balance to distance-covariance weighting, the effect of
source-to-target adaptation for existing continuous-treatment weighting
methods, and the additional effect of outcome augmentation.

The proposed estimator has the smallest MAB among the unaugmented
estimators for all source sample sizes. Its MAB decreases from 0.201 to
0.119 as \(n_S\) increases from 250 to 2000. The augmented proposed
estimator has MAB 0.178, 0.155, 0.137, and 0.120 across the same values
of \(n_S\). Outcome augmentation improves the proposed estimator most
noticeably when the source sample is small; for larger source samples,
the augmented and unaugmented versions are nearly indistinguishable.
This pattern is consistent with the weighting component becoming more
accurate as the labelled source sample grows.

The IRMSE results show the same pattern. The unaugmented proposed
estimator decreases from 0.261 at \(n_S=250\) to 0.145 at \(n_S=2000\),
and the augmented version decreases from 0.229 to 0.146. The augmented
proposed estimator has the smallest IRMSE for \(n_S=250,500,\) and
1000, while the unaugmented proposed estimator is slightly smaller at
\(n_S=2000\). Thus augmentation appears to provide finite-sample
stabilization, but it is not needed for the proposed weights to achieve
low estimation error when \(n_S\) is large.

The comparison with DCOW provides the most direct assessment of the
source-to-target extension. Unaugmented DCOW has MAB around 0.5 across
all source sample sizes, much larger than the corresponding proposed
estimator. This reflects the fact that DCOW removes treatment-covariate
dependence within the source sample but does not transport the source
covariate distribution to the target population. After outcome
augmentation, DCOW improves substantially and is competitive in MAB,
especially at \(n_S=250\). However, its IRMSE remains uniformly larger
than that of the proposed augmented estimator. This suggests that target
covariate balance is important for stabilizing the estimated curve, not
only for reducing average bias.

The adapted GPS methods provide a second comparison. Target adaptation
improves the unaugmented GPS estimators, particularly GPS-BART, showing
that correcting source-to-target covariate shift is beneficial even when
the continuous-treatment weights are estimated by generalized propensity
scores. Nevertheless, the adapted GPS estimators remain less accurate
than the proposed method. Outcome augmentation reduces MAB for several
GPS estimators, but it does not uniformly reduce IRMSE. In particular,
some augmented GPS estimators have moderate MAB but large IRMSE,
suggesting instability of the estimated curve across Monte Carlo
replications driven by interaction between misspecified nuisance models and the weighting procedure used.

Entropy balancing with outcome augmentation is the strongest
non-proposed competitor in several MAB comparisons, but its IRMSE remains
higher than that of the proposed estimator. This difference is consistent
with the distinction between balancing a finite set of moments and
directly targeting the distributional discrepancy. The adapted entropy
method improves upon its non-adapted version in some settings, but it
does not close the gap with the proposed source-to-target
distance-covariance criterion.

The supplementary tables show that increasing \(n_T\) has a smaller
effect than increasing \(n_S\) in this design. This agrees with the
structure of the problem: the target sample helps estimate the target
covariate distribution, whereas the labelled source sample drives both
confounding adjustment and outcome information. Taken together, the
simulation results indicate that directly balancing treatment-covariate
dependence and source-to-target covariate shift in one criterion gives
more accurate target ADRF estimates than source-only weighting or
two-step adaptation.

\begin{table}[ht!]
\centering
\caption{Simulation results for mean absolute bias (MAB), averaged over target sample sizes.}
\label{tab:simulation-mab-main}
\footnotesize
\setlength{\tabcolsep}{10pt}
\renewcommand{\arraystretch}{1.15}

\begin{tabular}{lcccccccc}
\toprule
& \multicolumn{4}{c}{Without augmentation}
& \multicolumn{4}{c}{With augmentation (DR version)} \\
\cmidrule(lr){2-5} \cmidrule(lr){6-9}
Method family
& $n_S=250$ & $500$ & $1000$ & $2000$
& $n_S=250$ & $500$ & $1000$ & $2000$ \\
\midrule
Proposed
& 0.201 & 0.165 & 0.139 & \textbf{0.119}
& 0.178 & \textbf{0.155} & \textbf{0.137} & 0.120 \\

DCOW
& 0.519 & 0.518 & 0.515 & 0.503
& \textbf{0.173} & 0.165 & 0.163 & 0.162 \\

GPS-BART
& 0.617 & 0.562 & 0.528 & 0.499
& 0.416 & 0.709 & 0.892 & 1.052 \\

GPS-BART-adapt
& 0.476 & 0.438 & 0.380 & 0.323
& 0.436 & 0.720 & 0.843 & 1.015 \\

GPS-LM
& 1.068 & 1.169 & 1.272 & 1.351
& 0.361 & 0.435 & 0.542 & 0.605 \\

GPS-LM-adapt
& 0.641 & 0.702 & 0.723 & 0.746
& 0.483 & 0.559 & 0.580 & 0.563 \\

Entropy
& 0.820 & 0.865 & 0.884 & 0.890
& 0.236 & 0.252 & 0.272 & 0.279 \\

Entropy-adapt
& 0.622 & 0.672 & 0.657 & 0.665
& 0.240 & 0.268 & 0.298 & 0.312 \\

Unweighted
& 0.827 & 0.825 & 0.830 & 0.824
& 0.340 & 0.343 & 0.353 & 0.354 \\
\bottomrule
\end{tabular}

\vspace{2pt}
\begin{minipage}{0.95\textwidth}
\footnotesize
\emph{Notes:} Entries are MAB averaged over
$n_T\in\{250,500,1000\}$ within each source sample size. The left block
reports the original estimators, and the right block reports the
corresponding outcome-augmented estimators. Bold entries indicate the
smallest rounded mean within each $n_S$ setting across all methods. Full
results stratified by both $n_S$ and $n_T$ are reported in Supplementary
Tables~S.1--S.4.
\end{minipage}
\end{table}

\begin{table}[ht!]
\centering
\caption{Simulation results for integrated root mean squared error (IRMSE), averaged over target sample sizes.}
\label{tab:simulation-irmse-main}
\footnotesize
\setlength{\tabcolsep}{10pt}
\renewcommand{\arraystretch}{1.15}

\begin{tabular}{lcccccccc}
\toprule
& \multicolumn{4}{c}{Without augmentation}
& \multicolumn{4}{c}{With augmentation (DR version)} \\
\cmidrule(lr){2-5} \cmidrule(lr){6-9}
Method family
& $n_S=250$ & $500$ & $1000$ & $2000$
& $n_S=250$ & $500$ & $1000$ & $2000$ \\
\midrule
Proposed
& 0.261 & 0.205 & 0.168 & \textbf{0.145}
& \textbf{0.229} & \textbf{0.192} & \textbf{0.165} & 0.146 \\

DCOW
& 0.547 & 0.531 & 0.521 & 0.507
& 0.242 & 0.212 & 0.193 & 0.183 \\

GPS-BART
& 0.681 & 0.618 & 0.577 & 0.541
& 1.212 & 1.554 & 1.626 & 1.673 \\

GPS-BART-adapt
& 0.690 & 0.655 & 0.584 & 0.511
& 1.727 & 2.070 & 2.176 & 1.859 \\

GPS-LM
& 1.202 & 1.288 & 1.364 & 1.440
& 1.297 & 1.745 & 1.933 & 2.004 \\

GPS-LM-adapt
& 0.812 & 0.865 & 0.853 & 0.859
& 1.503 & 1.892 & 2.110 & 1.930 \\

Entropy
& 0.907 & 0.892 & 0.908 & 0.899
& 0.367 & 0.314 & 0.308 & 0.305 \\

Entropy-adapt
& 1.217 & 0.923 & 0.728 & 0.720
& 0.393 & 0.351 & 0.341 & 0.345 \\

Unweighted
& 0.849 & 0.837 & 0.835 & 0.827
& 0.403 & 0.385 & 0.377 & 0.371 \\
\bottomrule
\end{tabular}

\vspace{2pt}
\begin{minipage}{0.95\textwidth}
\footnotesize
\emph{Notes:} Entries are IRMSE averaged over
$n_T\in\{250,500,1000\}$ within each source sample size. The left block
reports the original estimators, and the right block reports the
corresponding outcome-augmented estimators. Bold entries indicate the
smallest rounded mean within each $n_S$ setting across all methods. Full
results stratified by both $n_S$ and $n_T$ are reported in Supplementary
Tables~S.1--S.4.
\end{minipage}
\end{table}

\section{Dose-response analysis of PM\textsubscript{2.5} and county heart-disease mortality}
\label{sec:air_pollution_application}

We apply the proposed method to a county-level analysis of fine
particulate matter, PM\textsubscript{2.5}, and subsequent heart-disease
mortality. The baseline data are the EPA county-level PM\textsubscript{2.5},
cardiovascular mortality, and socioeconomic data for 2,132 US counties
over 1990--2010
\citep{epaPM25CMRData,wyatt2020annual,wyatt2020contribution}. The
treatment is the county-level mean predicted PM\textsubscript{2.5}
concentration averaged over 2011--2012, obtained from the CDC National
Environmental Public Health Tracking Network county PM\textsubscript{2.5}
files \citep{cdcPM25County2019}. The outcome is the CDC county
heart-disease mortality rate per 100,000 people, restricted to
county-level records with overall sex and overall race/ethnicity strata
\citep{cdcHeartMortalityCounty}. We use the three-year 2013--2015
mortality rate as the subsequent outcome. Further details on FIPS
harmonisation, variable construction, missing-value handling, and data
source identifiers are given in Supplementary
Section~F.

\begin{figure}[ht!]
    \centering
    \includegraphics[width=1\linewidth]{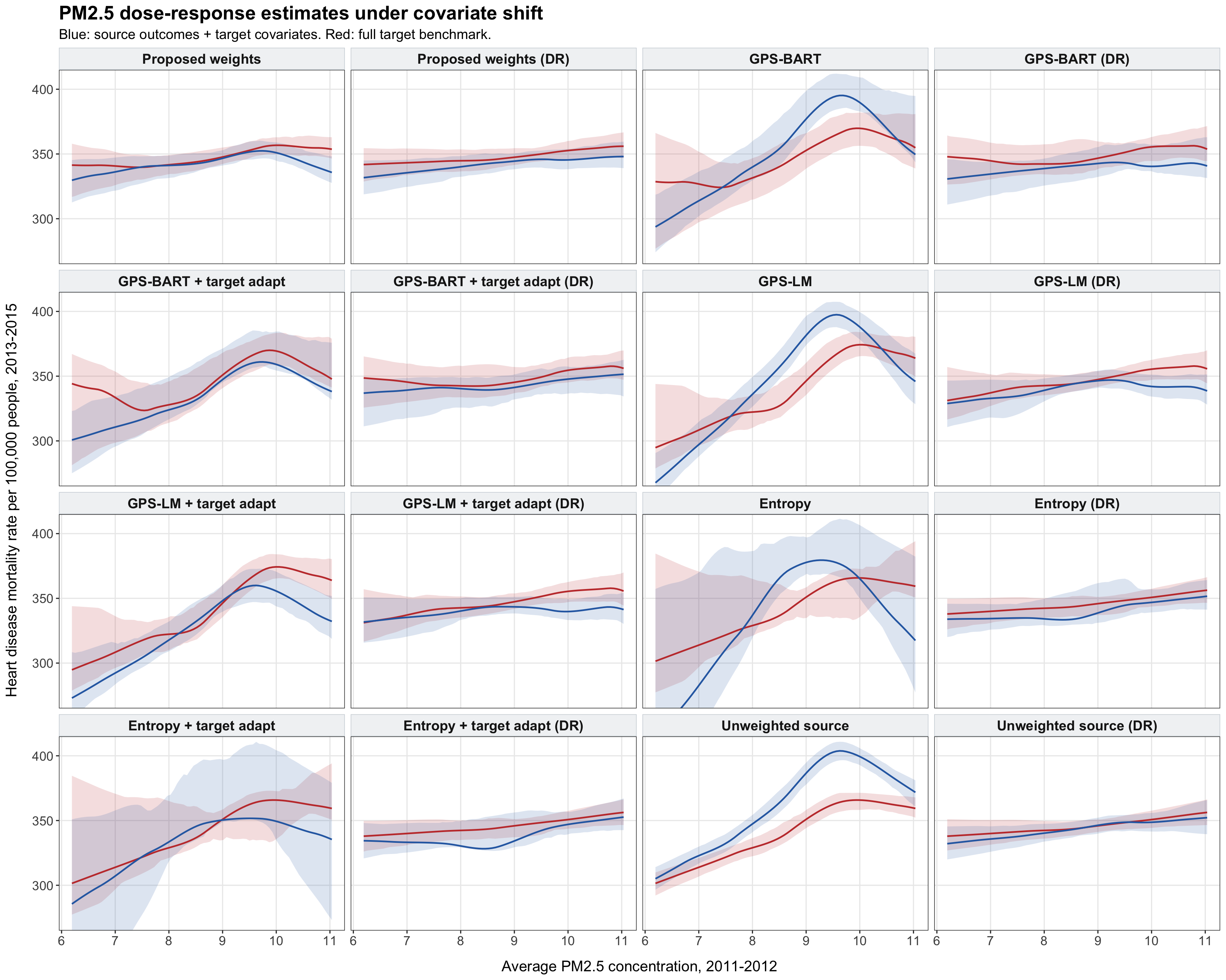}
    \caption{Dose-response estimates for PM\textsubscript{2.5} exposure
    and county heart-disease mortality under a source-target validation design. Blue curves use source outcomes and target covariates; red
    curves are empirical full-target benchmarks computed using target
    outcomes. Shaded bands are pointwise non-parametric bootstrap
    intervals.}
    \label{fig:air_pol_with_CI}
\end{figure}

This county-level PM\textsubscript{2.5} setting is related to the example
of \citet{bahadori2022end}, but here it is constructed as a
source-target transport analysis with temporally ordered covariates,
exposure, and outcome. To construct the validation design, we split
counties into source and target samples using a logistic sampling rule
based only on pre-treatment county characteristics. The source sample is
treated as labelled, so that \((X_i,A_i,Y_i)\) are observed. The target
sample is treated as unlabelled for the transported estimators, so that
only target covariates are used in estimation. Target treatment and
outcome values are held out from estimation and are used only to
construct empirical full-target benchmarks in
Figure~\ref{fig:air_pol_with_CI}. These benchmarks should not be
interpreted as the true ADRF; rather, they serve as a test of ``internal consistency'': they show the curves that would be
predicted if target outcomes were available.

For each method, we estimate the curve on the common support of source
and target PM\textsubscript{2.5}, using the interval between the larger
of the two fifth percentiles and the smaller of the two ninety-fifth
percentiles. Pointwise intervals are obtained by a non-parametric
bootstrap that resamples source counties and target counties separately.
For the adaptive GPS and entropy methods, the full-target benchmark uses
the corresponding non-adaptive within-target version, since no
source-to-target covariate-density correction is required when the
analysis is performed entirely in the target sample.

Figure~\ref{fig:air_pol_with_CI} shows the estimated dose-response
curves. The proposed transported estimates closely track the empirical
full-target benchmarks over the displayed PM\textsubscript{2.5} range.
The augmented proposed estimator is particularly stable, with a curve
that remains close and nearly parallel to the benchmark. Several
comparison methods show larger discrepancies. The non-augmented
unweighted and GPS-based estimators tend to produce more pronounced
high-exposure curvature, while their full-target benchmarks are smoother.
Target adaptation improves some GPS and entropy estimates, and outcome
augmentation generally produces smoother curves, but the proposed
transported estimates remain among the closest to the empirical
benchmarks. These results are consistent with the simulation findings:
directly balancing treatment-covariate dependence and source-to-target
covariate shift gives a stable estimate of the target ADRF.

The analysis is observational and ecological. A causal interpretation
requires the assumptions stated earlier, including no unmeasured
county-level confounding, positivity over the plotted PM\textsubscript{2.5}
range, and consistency of the county-level exposure and outcome
definitions. The source-target split is used as a validation design for
studying transport under covariate shift and is not intended to represent
a specific policy intervention or a naturally occurring sampling
mechanism.

\FloatBarrier

\section{Discussion}

We have developed a weighting-based framework for estimating average
dose-response functions under covariate shift. The proposed method is
motivated by the distributional target arising from the two-sample
pseudo-outcome: the weighted source distribution should remove
treatment-covariate dependence while also matching the target covariate
distribution. This leads to a source-to-target extension of
distance-covariance weighting for continuous treatments. We also develop
two-sample local polynomial regression for estimating the target ADRF
from labelled source data and unlabelled target covariates. The
theoretical analysis establishes uniform convergence of the proposed
optimization weights to the oracle source-to-target weights and derives
large-sample properties of the resulting TSLP estimator. These
weight-level results may also be useful for studying other
optimization-based causal weighting procedures.

The numerical studies show that the proposed weighting criterion improves
estimation of the target ADRF relative to source-only weighting,
generalized-propensity-score weighting, entropy balancing, and two-step
adaptation strategies. The comparison with DCOW indicates that removing
treatment-covariate dependence alone is not sufficient when the source
and target covariate distributions differ. The PM\textsubscript{2.5}
application gives a complementary empirical illustration: in a
source-target validation design, the proposed transported estimates track
the empirical full-target benchmark more closely than several competing
methods.

Several limitations suggest directions for future work. First, the
uniform convergence theorem is stated for approximate minimizers of the
empirical criterion, but obtaining quantitative rates for this
convergence remains open. Such rates may be possible by adapting
quantitative epigraphical-distance arguments such as those in
\citet{attouch1991quantitative}. The weight convergence result is stated
for approximate minimizers in an admissible function class; further work
is needed to connect this theory more directly to specific nonconvex
finite-sample solvers. Secondly, as in most causal analyses
based on observational data, identification relies on no unmeasured
confounding, positivity, and transportability of the conditional outcome
model. Extensions that allow for unmeasured confounding, for example
through instrumental-variable approaches or proximal causal learning
\citep{tchetgen2020introduction}, would be valuable. Finally, although
we have focused on the target ADRF, the same source-to-target weighting
principle may be useful for other continuous-treatment estimands and for
other smoothing or semiparametric estimators.

\label{sec:conc}

\bigskip
\begin{center}
{\large\bf SUPPLEMENTARY MATERIAL}
\end{center}

The supplementary material contains proofs of the theoretical results,
details on the source-to-target weighting criterion and its optimization,
additional implementation details for the simulation study, full simulation
results stratified by source and target sample sizes, and preprocessing
details for the PM\textsubscript{2.5} application.

\def\spacingset#1{\renewcommand{\baselinestretch}{#1}\small\normalsize} \spacingset{1.2}

\bibliographystyle{chicago}
\bibliography{bibfile}

@article{sen1974weak,
  title={Weak convergence of generalized U-statistics},
  author={Sen, Pranab Kumar},
  journal={The Annals of Probability},
  pages={90--102},
  year={1974},
  publisher={JSTOR}
}

@article{holford1981understanding,
  title={Understanding the dose-effect relationship: clinical application of pharmacokinetic-pharmacodynamic models},
  author={Holford, Nicholas HG and Sheiner, Lewis B},
  journal={Clinical pharmacokinetics},
  volume={6},
  number={6},
  pages={429--453},
  year={1981},
  publisher={Springer}
}

@article{imai2004causal,
  title={Causal inference with general treatment regimes: Generalizing the propensity score},
  author={Imai, Kosuke and Van Dyk, David A},
  journal={Journal of the American Statistical Association},
  volume={99},
  number={467},
  pages={854--866},
  year={2004},
  publisher={Taylor \& Francis}
}

@article{ramagopalan2022transportability,
  title={Transportability of overall survival estimates from US to Canadian patients with advanced non--small cell lung cancer with implications for regulatory and health technology assessment},
  author={Ramagopalan, Sreeram V and Popat, Sanjay and Gupta, Alind and Boyne, Devon J and Lockhart, Alexandre and Hsu, Grace and O’Sullivan, Dylan E and Inskip, Jessica and Ray, Joshua and Cheung, Winson Y and others},
  journal={JAMA Network Open},
  volume={5},
  number={11},
  pages={e2239874--e2239874},
  year={2022},
  publisher={American Medical Association}
}

@article{rubinstein2023balancing,
  title={Balancing weights for region-level analysis: the effect of medicaid expansion on the uninsurance rate among states that did not expand medicaid},
  author={Rubinstein, Max and Haviland, Amelia and Choi, David},
  journal={The Annals of Applied Statistics},
  volume={17},
  number={2},
  pages={1469--1490},
  year={2023},
  publisher={Institute of Mathematical Statistics}
}

@article{wu2024matching,
  title={Matching on generalized propensity scores with continuous exposures},
  author={Wu, Xiao and Mealli, Fabrizia and Kioumourtzoglou, Marianthi-Anna and Dominici, Francesca and Braun, Danielle},
  journal={Journal of the American Statistical Association},
  volume={119},
  number={545},
  pages={757--772},
  year={2024},
  publisher={Taylor \& Francis}
}

@article{rubin2005causal,
	title        = {Causal inference using potential outcomes: Design, modeling, decisions},
	author       = {Rubin, Donald B},
	year         = 2005,
	journal      = {J. Amer. Statist. Assoc.},
	publisher    = {Taylor \& Francis},
	volume       = 100,
	number       = 469,
	pages        = {322--331}
}

@article{huling2020energy,
	title = {Energy balancing of covariate distributions},
author = {Jared D. Huling and Simon Mak},
pages = {20220029},
volume = {12},
number = {1},
journal = {Journal of Causal Inference},
year = {2024},
}

@article{kallus2018confounding,
	title        = {Confounding-robust policy improvement},
	author       = {Kallus, Nathan and Zhou, Angela},
	year         = 2018,
	journal      = {Advances in neural information processing systems},
	volume       = 31
}

@article{imai2014covariate,
	title        = {Covariate balancing propensity score},
	author       = {Imai, Kosuke and Ratkovic, Marc},
	year         = 2014,
	journal      = {Journal of the Royal Statistical Society: Series B (Statistical Methodology)},
	publisher    = {Wiley Online Library},
	volume       = 76,
	number       = 1,
	pages        = {243--263}
}

@article{hainmueller2012entropy,
	title        = {Entropy balancing for causal effects: A multivariate reweighting method to produce balanced samples in observational studies},
	author       = {Hainmueller, Jens},
	year         = 2012,
	journal      = {Political analysis},
	publisher    = {Cambridge University Press},
	volume       = 20,
	number       = 1,
	pages        = {25--46}
}

@article{wang2020minimal,
	title        = {Minimal dispersion approximately balancing weights: asymptotic properties and practical considerations},
	author       = {Wang, Yixin and Zubizarreta, Jose R},
	year         = 2020,
	journal      = {Biometrika},
	publisher    = {Oxford University Press},
	volume       = 107,
	number       = 1,
	pages        = {93--105}
}

@article{chen2023robust,
  title={Robust Sample Weighting to Facilitate Individualized Treatment Rule Learning for a Target Population},
  author={Chen, Rui and Huling, Jared D and Chen, Guanhua and Yu, Menggang},
  journal={Biometrika},
  pages={asad038},
  year={2023},
  publisher={Oxford University Press}
}

@article{imbens2000role,
	title        = {The role of the propensity score in estimating dose-response functions},
	author       = {Imbens, Guido W},
	year         = 2000,
	journal      = {Biometrika},
	publisher    = {Oxford University Press},
	volume       = 87,
	number       = 3,
	pages        = {706--710}
}

@book{hirano2004propensity,
	title        = {The Propensity Score with Continuous Treatments},
	author       = {Hirano, Keisuke and Imbens, Guido W},
	year         = 2004,
	journal      = {Applied Bayesian Modeling and Causal Inference from Incomplete-Data Perspectives: An Essential Journey with Donald Rubin's Statistical Family},
	publisher    = {Wiley Online Library},
	pages        = {73--84}
}

@article{kennedy2017non,
	title        = {Non-parametric methods for doubly robust estimation of continuous treatment effects},
	author       = {Kennedy, Edward H and Ma, Zongming and McHugh, Matthew D and Small, Dylan S},
	year         = 2017,
	journal      = {Journal of the Royal Statistical Society Series B: Statistical Methodology},
	publisher    = {Oxford University Press},
	volume       = 79,
	number       = 4,
	pages        = {1229--1245}
}

@article{takatsu2022debiased,
	title={Debiased inference for a covariate-adjusted regression function},
  author={Takatsu, Kenta and Westling, Ted},
  journal={Journal of the Royal Statistical Society Series B: Statistical Methodology},
  volume={87},
  number={1},
  pages={33--55},
  year={2025},
  publisher={Oxford University Press UK}
}

@article{kim2026estimating,
  title={Estimating Continuous Treatment Effects with Two-Stage Kernel Ridge Regression},
  author={Kim, Seok-Jin and Wang, Kaizheng},
  journal={arXiv preprint arXiv:2604.13410},
  year={2026}
}

@article{colangelo2020double,
	title        = {Double debiased machine learning nonparametric inference with continuous treatments},
	author       = {Colangelo, Kyle and Lee, Ying-Ying},
	year         = 2020,
	journal      = {arXiv preprint arXiv:2004.03036}
}

@article{huling2023independence,
	title        = {Independence weights for causal inference with continuous treatments},
	author       = {Huling, Jared D and Greifer, Noah and Chen, Guanhua},
	year         = 2024,
	journal      = {Journal of the American Statistical Association},
	publisher    = {Taylor \& Francis},
	volume       = 119,
	number       = 546,
	pages        = {1657--1670}
}

@article{szekely2007measuring,
  title={Measuring and testing dependence by correlation of distances},
  author={Sz{\'e}kely, G{\'a}bor J and Rizzo, Maria L and Bakirov, Nail K},
  journal={Annals of Statistics},
  volume={35},
  number={6},
  pages={2769--2794},
  year={2007},
  publisher={Institute of Mathematical Statistics}
}

@article{stellato2020osqp,
	title        = {OSQP: An operator splitting solver for quadratic programs},
	author       = {Stellato, Bartolomeo and Banjac, Goran and Goulart, Paul and Bemporad, Alberto and Boyd, Stephen},
	year         = 2020,
	journal      = {Mathematical Programming Computation},
	publisher    = {Springer},
	volume       = 12,
	number       = 4,
	pages        = {637--672}
}

@article{singh2020kernel,
	title={Kernel methods for causal functions: dose, heterogeneous and incremental response curves},
  author={Singh, Rahul and Xu, Liyuan and Gretton, Arthur},
  journal={Biometrika},
  volume={111},
  number={2},
  pages={497--516},
  year={2024},
  publisher={Oxford University Press}
}

@article{szekely2013energy,
	title        = {Energy statistics: A class of statistics based on distances},
	author       = {Sz{\'e}kely, G{\'a}bor J and Rizzo, Maria L},
	year         = 2013,
	journal      = {Journal of statistical planning and inference},
	publisher    = {Elsevier},
	volume       = 143,
	number       = 8,
	pages        = {1249--1272}
}

@book{lee2019u,
	title        = {U-statistics: Theory and Practice},
	author       = {Lee, A J},
	year         = 2019,
	publisher    = {Routledge}
}

@article{vegetabile2021nonparametric,
	title        = {Nonparametric estimation of population average dose-response curves using entropy balancing weights for continuous exposures},
	author       = {Vegetabile, Brian G and Griffin, Beth Ann and Coffman, Donna L and Cefalu, Matthew and Robbins, Michael W and McCaffrey, Daniel F},
	year         = 2021,
	journal      = {Health Services and Outcomes Research Methodology},
	publisher    = {Springer},
	volume       = 21,
	pages        = {69--110}
}

@article{tubbicke2021entropy,
	title        = {Entropy balancing for continuous treatments},
	author       = {T{\"u}bbicke, Stefan},
	year         = 2021,
	journal      = {Journal of Econometric Methods},
	publisher    = {De Gruyter},
	volume       = 11,
	number       = 1,
	pages        = {71--89}
}

@article{attouch1991quantitative,
	title        = {Quantitative stability of variational systems. I. The epigraphical distance},
	author       = {Attouch, H{\'e}dy and Wets, Roger J-B},
	year         = 1991,
	journal      = {Transactions of the American Mathematical Society},
	volume       = 328,
	number       = 2,
	pages        = {695--729}
}

@article{gong2019higher,
  title={Higher US rural mortality rates linked to socioeconomic status, physician shortages, and lack of health insurance},
  author={Gong, Gordon and Phillips, Scott G and Hudson, Catherine and Curti, Debra and Philips, Billy U},
  journal={Health Affairs},
  volume={38},
  number={12},
  pages={2003--2010},
  year={2019}
}

@article{carre2017does,
  title={Does air pollution play a role in infertility?: a systematic review},
  author={Carr{\'e}, Julie and Gatimel, Nicolas and Moreau, Jessika and Parinaud, Jean and L{\'e}andri, Roger},
  journal={Environmental Health},
  volume={16},
  pages={1--16},
  year={2017},
  publisher={Springer}
}

@article{fong2018covariate,
  title={Covariate balancing propensity score for a continuous treatment: Application to the efficacy of political advertisements},
  author={Fong, Christian and Hazlett, Chad and Imai, Kosuke},
  journal={The Annals of Applied Statistics},
  volume={12},
  number={1},
  pages={156--177},
  year={2018},
  publisher={JSTOR}
}

@article{sugiyama2007direct,
  title={Direct importance estimation with model selection and its application to covariate shift adaptation},
  author={Sugiyama, Masashi and Nakajima, Shinichi and Kashima, Hisashi and Buenau, Paul and Kawanabe, Motoaki},
  journal={Advances in neural information processing systems},
  volume={20},
  year={2007}
}

@article{huang2006correcting,
  title={Correcting sample selection bias by unlabeled data},
  author={Huang, Jiayuan and Gretton, Arthur and Borgwardt, Karsten and Sch{\"o}lkopf, Bernhard and Smola, Alex},
  journal={Advances in neural information processing systems},
  volume={19},
  year={2006}
}

@article{degtiar2023review,
  title={A review of generalizability and transportability},
  author={Degtiar, Irina and Rose, Sherri},
  journal={Annual Review of Statistics and Its Application},
  volume={10},
  number={1},
  pages={501--524},
  year={2023},
  publisher={Annual Reviews}
}

@article{pan2009survey,
  title={A survey on transfer learning},
  author={Pan, Sinno Jialin and Yang, Qiang},
  journal={IEEE Transactions on knowledge and data engineering},
  volume={22},
  number={10},
  pages={1345--1359},
  year={2009},
  publisher={IEEE}
}

@article{tchetgen2020introduction,
  title={An introduction to proximal causal learning},
  author={Tchetgen, Eric J Tchetgen and Ying, Andrew and Cui, Yifan and Shi, Xu and Miao, Wang},
  journal={arXiv preprint arXiv:2009.10982},
  year={2020}
}

@inproceedings{bahadori2022end,
  title     = {End-to-End Balancing for Causal Continuous Treatment-Effect Estimation},
  author    = {Bahadori, Taha and Tchetgen Tchetgen, Eric and Heckerman, David},
  booktitle = {Proceedings of the 39th International Conference on Machine Learning},
  series    = {Proceedings of Machine Learning Research},
  volume    = {162},
  pages     = {1313--1326},
  year      = {2022},
  publisher = {PMLR},
  url       = {https://proceedings.mlr.press/v162/bahadori22a.html}
}

@misc{epaPM25CMRData,
  author       = {{U.S. Environmental Protection Agency Office of Research and Development}},
  title        = {Annual {PM2.5} and Cardiovascular Mortality Rate Data: Trends Modified by County Socioeconomic Status in 2,132 {US} Counties},
  year         = {2019},
  howpublished = {Data.gov dataset},
  doi          = {10.23719/1506014},
  url          = {https://catalog.data.gov/dataset/annual-pm2-5-and-cardiovascular-mortality-rate-data-trends-modified-by-county-socioeconomi},
  note         = {Accessed 4 May 2026}
}

@article{wyatt2020annual,
  title   = {Annual {PM2.5} and Cardiovascular Mortality Rate Data: Trends Modified by County Socioeconomic Status in 2,132 {US} Counties},
  author  = {Wyatt, Lauren H. and Peterson, Geoffrey Colin L. and Wade, Tim J. and Neas, Lucas M. and Rappold, Ana G.},
  journal = {Data in Brief},
  volume  = {30},
  pages   = {105318},
  year    = {2020},
  doi     = {10.1016/j.dib.2020.105318}
}

@article{wyatt2020contribution,
  title   = {The Contribution of Improved Air Quality to Reduced Cardiovascular Mortality: Declines in Socioeconomic Differences over Time},
  author  = {Wyatt, Lauren H. and Peterson, Geoffrey C. L. and Wade, Timothy J. and Neas, Lucas M. and Rappold, Ana G.},
  journal = {Environment International},
  volume  = {136},
  pages   = {105430},
  year    = {2020},
  doi     = {10.1016/j.envint.2019.105430}
}

@misc{cdcPM25County2019,
  author       = {{Centers for Disease Control and Prevention, National Environmental Public Health Tracking Network}},
  title        = {Daily County-Level {PM2.5} Concentrations, 2001--2019},
  year         = {2023},
  howpublished = {Data.CDC.gov dataset},
  url          = {https://data.cdc.gov/Environmental-Health-Toxicology/Daily-County-Level-PM2-5-Concentrations-2001-2019/dqwm-pbi7},
  note         = {Modeled county-level predictions from EPA's Downscaler model. Accessed 4 May 2026}
}

@misc{cdcHeartMortalityCounty,
  author       = {{Centers for Disease Control and Prevention, Division for Heart Disease and Stroke Prevention}},
  title        = {Heart Disease Mortality Data Among {US} Adults (35+) by State/Territory and County},
  year         = {2023},
  howpublished = {Data.CDC.gov dataset},
  url          = {https://data.cdc.gov/Heart-Disease-Stroke-Prevention/Heart-Disease-Mortality-Data-Among-US-Adults-35-by/48mw-5apu},
  note         = {Age-standardized, spatially smoothed, three-year average county rates. Accessed 4 May 2026}
}

@article{doss2024nonparametric,
  title={A nonparametric doubly robust test for a continuous treatment effect},
  author={Doss, Charles R and Weng, Guangwei and Wang, Lan and Moscovice, Ira and Chantarat, Tongtan},
  journal={The Annals of Statistics},
  volume={52},
  number={4},
  pages={1592--1615},
  year={2024},
  publisher={Institute of Mathematical Statistics}
}

\end{document}